\documentclass{article}
\usepackage{icmc2020template}
\usepackage{times}
\usepackage{ifpdf}
\usepackage{soul}
\usepackage{bm}
\usepackage[english]{babel}

\usepackage{amsmath} 
\usepackage{amssymb} 

\usepackage{enumitem}

\def\papertitle{Neural Granular Sound Synthesis}
\def\firstauthor{Adrien Bitton}
\def\secondauthor{Philippe Esling}
\def\thirdauthor{Tatsuya Harada}

\newif\ifpdf
\ifx\pdfoutput\relax
\else
   \ifcase\pdfoutput
      \pdffalse
   \else
      \pdftrue
  \fi
\fi

\ifpdf 
  \usepackage[pdftex,
    pdftitle={\papertitle},
    pdfauthor={\firstauthor, \secondauthor, \thirdauthor},
    bookmarksnumbered, 
    pdfstartview=XYZ 
   ]{hyperref}

  \usepackage[pdftex]{graphicx}
  \graphicspath{{./figures/}}
  \DeclareGraphicsExtensions{.pdf,.jpeg,.png}

  \usepackage[figure,table]{hypcap}

\else 
  \usepackage[dvips,
    bookmarksnumbered, 
    pdfstartview=XYZ 
  ]{hyperref}  

  \usepackage[dvips]{epsfig,graphicx}
  \graphicspath{{./figures/}}
  \DeclareGraphicsExtensions{.eps}

  \usepackage[figure,table]{hypcap}
\fi

\hypersetup{
    colorlinks,%
    citecolor=black,%
    filecolor=black,%
    linkcolor=black,%
    urlcolor=black
}

\usepackage{ctable}

\title{\papertitle}

 \threeauthors
   {0.5in}
   {\firstauthor} {IRCAM, CNRS UMR9912\\Paris, France \\ %
     {\tt \href{mailto:bitton@ircam.fr}{bitton@ircam.fr}}}
   {\secondauthor} {IRCAM, CNRS UMR9912\\Paris, France \\ %
     {\tt \href{mailto:esling@ircam.fr}{esling@ircam.fr}}}
   {\thirdauthor} {The University of Tokyo, RIKEN\\Tokyo, Japan \\ %
     {\tt \href{mailto:harada@mi.t.u-tokyo.ac.jp}{harada@mi.t.u-tokyo.ac.jp}}}

\begin{document}
\capstartfalse
\maketitle
\capstarttrue

\begin{abstract}
Granular sound synthesis is a popular audio generation technique based on rearranging sequences of small waveform windows. In order to control the synthesis, all grains in a given corpus are analyzed through a set of acoustic descriptors. This provides a representation reflecting some form of local similarities across the grains. However, the quality of this grain space is bound by that of the descriptors. Its traversal is not continuously invertible to signal and does not render any structured temporality.

We demonstrate that generative neural networks can implement granular synthesis while alleviating most of its shortcomings. We efficiently replace its audio descriptor basis by a probabilistic latent space learned with a Variational Auto-Encoder. In this setting the learned grain space is invertible, meaning that we can continuously synthesize sound when traversing its dimensions. It also implies that original grains are not stored for synthesis. Another major advantage of our approach is to learn structured paths inside this latent space by training a higher-level temporal embedding over arranged grain sequences.

The model can be applied to many types of libraries, including pitched notes or unpitched drums and environmental noises. We report experiments on the common granular synthesis processes as well as novel ones such as conditional sampling and morphing.
\end{abstract}

\section{Introduction}\label{sec:intro}
The process of generating musical audio has seen a continuous expansion since the advent of digital systems. Audio synthesis methods relying on parametric models can be derived from physical considerations, spectral analysis (e.g. sinusoids plus noise \cite{sms} models) or signal processing operations (e.g. frequency modulation). Alternatively to those signal generation techniques, samplers provide synthesis mechanisms by relying on stored waveforms and sets of audio transformations. However, when tackling large audio sample libraries, these methods cannot scale and are also unable to aggregate a model over the whole data. Therefore, they cannot globally manipulate the audio features in the sound generation process. To this extent, corpus-based synthesis has been introduced by slicing sets of signals in shorter audio segments, which can be rearranged into new waveforms through a selection algorithm.
\begin{figure}[ht]
\centering
\includegraphics[width=0.95\columnwidth]{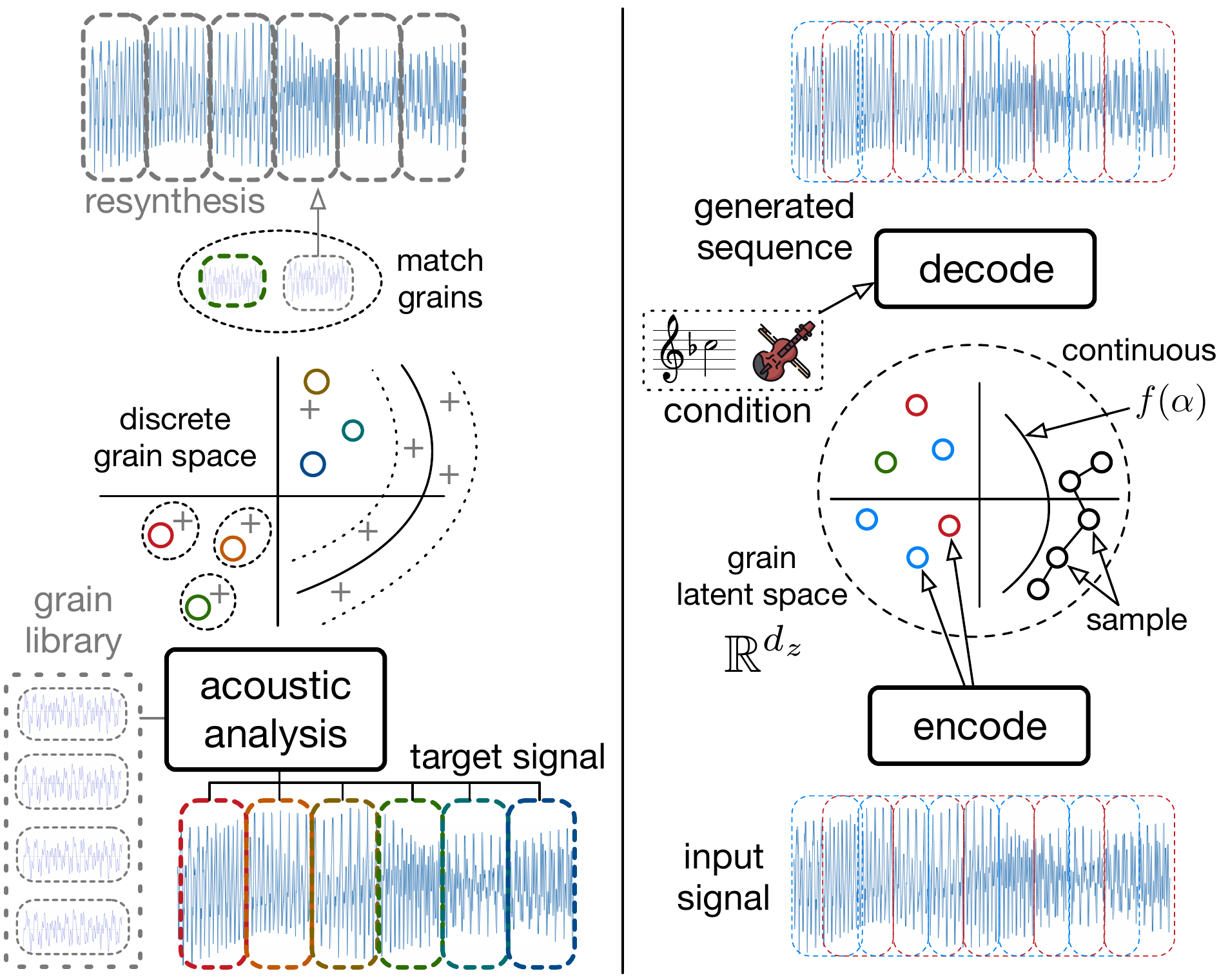}
\caption{Left: A grain library is analysed and scattered (+) into the acoustic dimensions. A target is defined, by analysing an other signal (o) or as a free trajectory, and matched to the library through the acoustic descriptors. Subsequently, grains are selected and arranged into a waveform. Right: The grain latent space can continuously synthesize waveform grains. Latent features can be encoded from an input signal, sampled from a structured temporal embedding or freely drawn. Explicit controls can be learned as target conditions for the decoder.\label{fig:neur_gran}}
\end{figure}

An instance of corpus-based synthesis, named \textit{granular sound synthesis} \cite{Roads_GS}, uses short waveform windows of a fixed length. These units (called \textit{grains}) usually have a size ranging between 10 and 100 milliseconds. For a given corpus, the grains are extracted and can be analyzed through audio descriptors \cite{adesc} in order to facilitate their manipulation. Such analysis space provides a representation that reflects some form of local similarities across grains. The grain corpus is displayed as a cloud of points whose distances relate to some of their acoustic relationships. By relying on this space, resynthesis can be done with \textit{concatenative sound synthesis} \cite{catart}. To a certain extent, this process can emulate the spectro-temporal dynamics of a given signal. However, the perceptual quality of the audio similarities, assessed through predefined sets of acoustic descriptors, is inherently biased by their design. These only offer a limited consistency across many different sounds, within the corpus and with respect to other targets. Furthermore, it should be noted that the synthesis process can only use the original grains, precluding continuously invertible interpolations in this grain space.

To enhance the expressivity of granular synthesis, grain sequences should be drawn in more flexible ways, by understanding the temporal dynamics of trajectories in the acoustic descriptor space. However, current methods are only restricted to perform random or simple hand-drawn paths. Traversals across the space map to grain series that are ordered according to the corresponding features. However, given that the grain space from current approaches is not invertible, these paths do not correspond to continuous audio synthesis, besides that of each of the scattered original grains. This could be alleviated by having a denser grain space (leading to a smoother assembled waveform), but it would require a correspondingly increasing amount of memory, quickly exceeding the gigabyte scale when considering nowadays sound sample library sizes. In a real-time setting, this causes further limitations to consider in a traditional granular synthesis space. As current methods only account for local relationships, they cannot generate the structured temporal dynamics of musical notes or drum hits without having a strong inductive bias, such as a target signal. Finally, the audio descriptors and the slicing size of grains are critical parameters to choose for these methods. They model the perceptual relationships across elements and set a trade-off: shorter grains allow for a denser space and faster sound variations at the expense of a limited estimate of the spectral features and the need to process larger series for a given signal duration. 

In this paper, we show that we can address most of the aforementioned shortcomings by drawing parallels between granular sound synthesis and probabilistic latent variable models. We develop a new neural granular synthesis technique that refines granular synthesis and is efficiently solved by generative neural networks (Figure \ref{fig:neur_gran}). Through the repeated observation of grains, our proposed technique adaptively and unsupervisedly learns analysis dimensions, structuring a latent grain space, which is continuously invertible to signal domain. Such space embeds the training dataset, which is no longer required in memory for generation. It allows to continuously generate novel grains at any interpolated latent position. In a second step, this space serves as basis for a higher-level temporal modeling, by training a sequential embedding over contiguous series of grain features. As a result, we can sample latent paths with a consistent temporal structure and moreover relieve some of the challenges to learn to generate raw waveforms. Its architecture is suited to optimizing local spectro-temporal features that are essential for audio quality, as well as longer-term dependencies that are efficiently extracted from grain-level sequences rather than individual waveform samples. The trainable modules used are well-grounded in digital signal processing (DSP), thus interpretable and efficient for sound synthesis. By providing simple variations of the model, it can adapt to many audio domains as well as different user interactions. With this motivation, we report several experiments applying the creative potentials of granular synthesis to neural waveform modeling: continuous free-synthesis with variable step size, one-shot sample generation with controllable attributes, analysis/resynthesis for audio style transfer and high-level interpolation between audio samples.

\section{State of the art}\label{sec:sota}

\subsection{Generative neural networks}\label{subsec:generative_nn}

\textit{Generative models} aim to understand a given set $\mathbf{x}\in\mathbb{R}^{d_{x}}$ by modeling an underlying probability distribution $p(\mathbf{x})$ of the data. To do so, we consider \emph{latent variables} defined in a lower-dimensional space $\mathbf{z}\in\mathbb{R}^{d_{z}}$ ($d_{z} \ll d_{x}$), as a higher-level representation \textit{generating} any given example. The complete model is defined by $p(\mathbf{x}, \mathbf{z}) = p(\mathbf{x} \vert \mathbf{z})p(\mathbf{z})$. However, a real-world dataset follows a complex distribution that cannot be evaluated analytically. The idea of \emph{variational inference} (VI) is to address this problem through \emph{optimization} by assuming a simpler distribution $q_{\phi}(\mathbf{z}\vert\mathbf{x})\in\mathcal{Q}$ from a family of approximate densities \cite{vae}. The goal of VI is to minimize differences between the approximated and real distribution, by using their Kullback-Leibler (KL) divergence
\begin{equation}
q_{\phi}^{*}(\mathbf{z}\vert \mathbf{x})=\underset{q_{\phi}(\mathbf{z} \vert \mathbf{x})\in\mathcal{Q}}{\text{argmin}} \mathcal{D}_{KL} \big[ q_{\phi}\left(\mathbf{z} \vert \mathbf{x}\right) \parallel p_\theta\left(\mathbf{z} \vert \mathbf{x}\right) \big].
\end{equation}
By developing this divergence and re-arranging terms (detailed development can be found in \cite{vae}), we obtain
\begin{multline}
    \log{p(\mathbf{x})} - \mathcal{D}_{KL} \big[ q_{\phi}(\mathbf{z} \vert \mathbf{x}) \parallel p_{\theta}(\mathbf{z} \vert \mathbf{x}) \big] \\ 
= \mathbb{E}_{\mathbf{z}} \big[ \log{p(\mathbf{x} \vert \mathbf{z})}\big] - \mathcal{D}_{KL} \big[ q_{\phi}(\mathbf{z} \vert \mathbf{x}) \parallel p_{\theta}(\mathbf{z}) \big].
\end{multline}
This formulation of the \textit{Variational Auto-Encoder} (VAE) relies on an encoder $q_{\phi}(\mathbf{z}|\mathbf{x})$, which aims at minimizing the distance to the unknown conditional latent distribution. Under this assumption, the Evidence Lower Bound Objective (ELBO) is optimized by minimization of a $\beta$ weighted KL regularization over the latent distribution added to the reconstruction cost of the decoder $p_\theta (\mathbf{x|z})$
\begin{equation}
\resizebox{.88\hsize}{!}{$\mathcal{L}_{\theta,\phi}=\underbrace{-\mathbb{E}_{q_\phi(\mathbf{z})}\big[ \log{ p_\theta (\mathbf{x|z}) } \big]}_{\text{reconstruction}}+\beta*\underbrace{\mathcal{D}_{KL} \big[ q_\phi(\mathbf{z|x}) \parallel p_\theta(\mathbf{z}) \big]}_{\text{regularization}}$}.
\label{eq:ELBO}
\end{equation}
The second term of this loss requires to define a prior distribution over the latent space, which for ease of sampling and back-propagation is chosen to be an isotropic gaussian of unit variance $p_{\theta}(\mathbf{z})=\mathcal{N}(\mathbf{0},\mathbf{I})$. Accordingly, a forward pass of the VAE consists in \textit{encoding} a given data point $q_\phi : \mathbf{x}\xrightarrow{}\{\bm{\mu}(\mathbf{x}),\bm{\sigma}(\mathbf{x})\}$ to obtain a mean $\bm{\mu}(\mathbf{x})$ and variance $\bm{\sigma}(\mathbf{x})$. These allow us to obtain the latent $\mathbf{z}$ by sampling from the Gaussian, such that $\mathbf{z}\sim\mathcal{N}(\bm{\mu}(\mathbf{x}),\bm{\sigma}(\mathbf{x}))$.

The representation learned with a VAE has a smooth topology \cite{higgins2016beta} since its encoder is regularized on a continuous density and intrinsically supports sampling within its unsupervised training process. Its latent dimensions can serve both for analysis when encoding new samples, or as generative variables that can continuously be decoded back to the target data domain. Furthermore, it has been shown \cite{esling2018generative} that it could be successfully applied to audio generation. Thus, it is the core of our neural model for granular synthesis of raw waveforms.

\subsection{Neural waveform generation}\label{subsec:neural_waveform}

Applications of generative neural networks to raw audio data must face the challenge of modeling time series with very high sampling rates. Hence, the models must account for both local features ensuring the generated audio quality, as well as longer-term relationships (consistent over tens of thousands of samples) in order to form meaningful signals. The first proposed approaches were based on auto-regressive models, which exploit the causal nature of audio. Given the whole waveform $\mathbf{x}=\{x_1,\ldots,x_T\}$, these models decompose the joint distribution into a product of conditional distributions. Hence, each sample is generated conditionally on all previous ones
\begin{equation}
p(\mathbf{x})=\prod_{t=1}^{T}p(x_t|x_1,\ldots,x_{t-1}).
\label{eq:wavenet}
\end{equation}
Amongst these models, WaveNet \cite{wavenet} has been established as the reference solution for high-quality speech synthesis. It has also been successfully applied to musical audio with the Nsynth dataset \cite{nsynth}. However, generating a signal in an auto-regressive manner is inherently slow since it iterates one sample at a time. Moreover, a large convolutional structure is needed in order to infer even a limited context of 100ms. This results in heavy models, only adapted to large databases and requiring long training times.

Specifically for musical audio generation, the Symbol-to-Instrument Neural Generator (SING) proposes an overlap-add convolutional architecture \cite{sing} on top of which a sequential embedding $S$ is trained on frame steps $\mathbf{F}_{1\ldots f}$, by conditioning over instrument, pitch and velocity classes $(\mathbf{I},\mathbf{P},\mathbf{V})$. The model processes signal windows of 1024 points with a 75\% overlap, thus reducing the temporal dimension by 256 before the forward pass of the up-sampling convolutional decoder $D$. Given an input signal with log-magnitude spectrogram $l(\mathbf{x})=\log(\epsilon+|\text{STFT}[\mathbf{x}]|^2)$, the decoder outputs a reconstruction $\hat{\mathbf{x}}$, in order to optimize
\begin{equation}
\underset{D,S}{\text{argmin }}||l(\mathbf{x}) - l(\hat{\mathbf{x}})||_1
\label{eq:SING}
\end{equation}
for $\hat{\mathbf{x}}=D(S(\mathbf{F},\mathbf{I},\mathbf{P},\mathbf{V}))$. This approach removes  auto-regressive computation costs and offers meaningful controls, while achieving high-quality synthesis. However, given its specific architecture, it does not generalize to generative tasks other than sampling individual instrumental notes of fixed duration in pitched domains.

Recently, additional inductive biases arising from digital signal processing have allowed to specify tighter constraints on model definitions, leading to high sound quality with lower training costs. In this spirit, the Neural Source-Filter (NSF) model \cite{nsf} applies the idea of Spectral Modeling Synthesis (SMS) \cite{sms} to speech synthesis. Its input module receives acoustic features and computes conditioning information for the source and temporal filtering modules. In order to render both voiced and unvoiced sounds, a sinusoidal and gaussian noise excitations are fed into separate filter modules. 
Estimation of noisy and harmonic components is further improved by relying on a multi-scale spectrogram reconstruction criterion. 

Similar to NSF, but for pitched musical audio, the Differentiable Digital Signal Processing \cite{ddsp} model has been proposed. Compared to NSF, this architecture features an harmonic additive synthesizer that is summed with a subtractive noise synthesizer. Envelopes for the fundamental frequency and loudness as well as latent features are extracted from a waveform and fed into a recurrent decoder which controls both synthesizers. 
An alternative filter design is proposed by learning frequency-domain transfer functions of time-varying Finite Impulse Response (FIR) filters. 
Furthermore, the summed output is fed into a reverberation module that refines the acoustic quality of the signal. Although this process offers very promising results, it is restricted in the nature of signals that can be generated. 

\section{Neural granular sound synthesis}\label{sec:model_granular}

\begin{figure}[ht]
\centering
\includegraphics[width=1.\columnwidth]{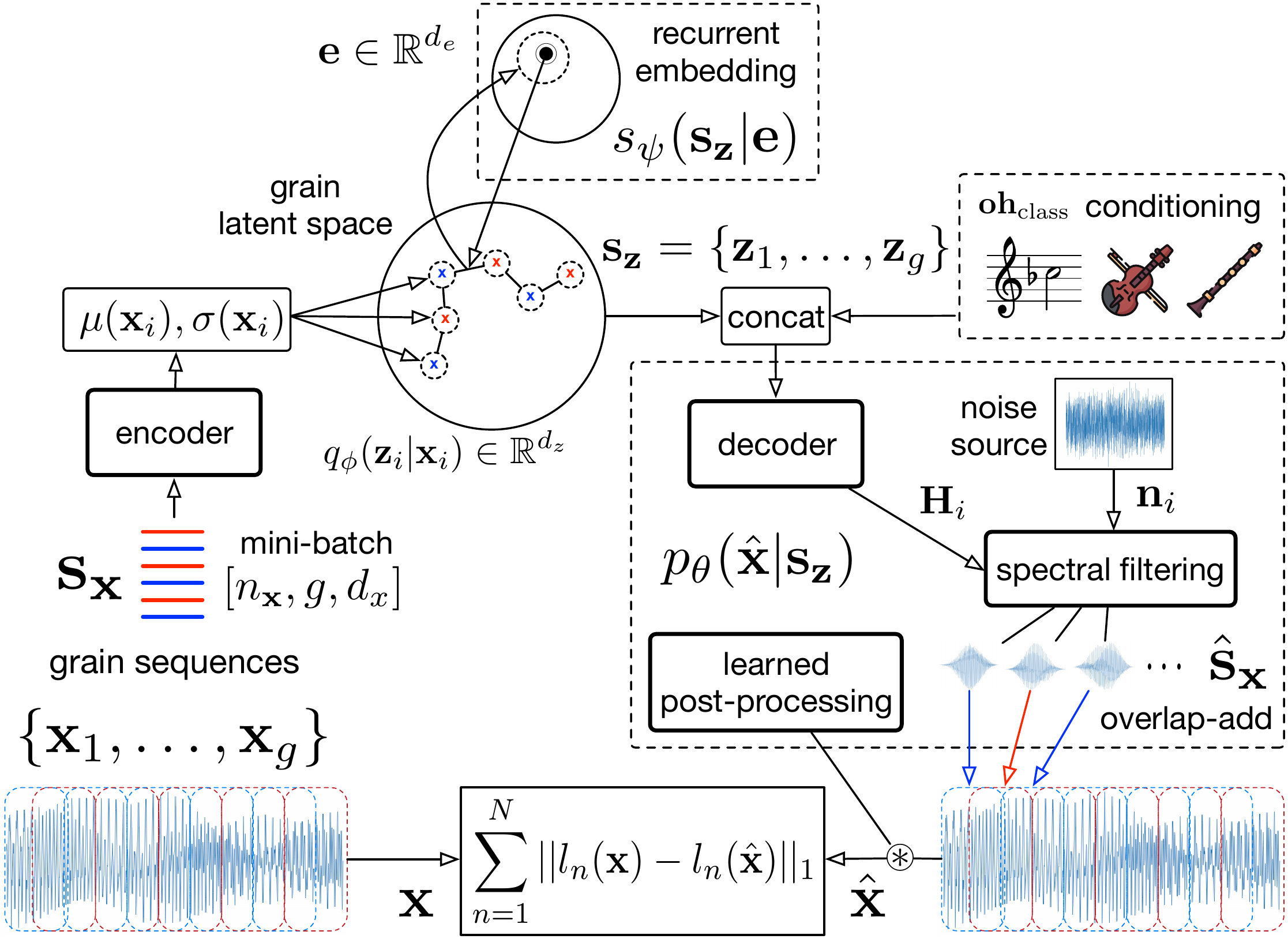}
\caption{Overview of the neural granular sound synthesis model.\label{fig:architecture}}
\end{figure}

In this paper, we propose a model that can learn both a local audio representation and modeling at multiple time scales, by introducing a neural version of the \textit{granular sound synthesis} \cite{catart}. The audio quality of short-term signal windows is ensured by efficient DSP modules optimized with a spectro-temporal criterion suited to both periodic and stochastic components. We structure the relative acoustic relationships in a latent grain space, by explicitly reconstructing waveforms through an \textit{overlap-add mechanism} across audio grain sequences. This synthesis operation can model any type of spectrogram, while remaining interpretable. Our proposal allows for analysis prior to data-driven resynthesis and also performs continuous variable length free-synthesis trajectories. Taking advantage of this grain-level representation, we further train a higher-level sequence embedding to generate audio events with meaningful temporal structure. In its less restrictive definition, our model allows for unconditional sampling, but it can be trained with additional independent controls (such as pitch or user classes) for more explicit interactions in composition and sound transfer. The complete architecture is depicted in Figure \ref{fig:architecture}.

\subsection{Latent grain space}\label{subsec:latent_grain}

Formally, we consider a set $\mathcal{X}$ of audio grains $\mathbf{x}_{i}\in\mathbb{R}^{d_{x}}$ extracted from audio waveforms $\mathbf{x}$ in a given sound corpus, with fixed grain size $d_{x}$. This set of grains follows an underlying probability density $p(\mathbf{x}_{i})$ that we aim to approximate through a parametric distribution $p_\theta$. This would allow to synthesize consistent novel audio grains by sampling $\hat{\mathbf{x}}_{j}\sim p_\theta(\mathbf{x}_{i})$. This likelihood is usually intractable, we can tackle this process by introducing a set of latent variables $\mathbf{z}\in\mathbb{R}^{d_{z}}$ ($d_{z} \ll d_{x}$). This low-dimensional space is expected to represent the most salient features of the data, which might have led to generate a given example. In our case, it will efficiently replace the use of acoustic descriptors, by optimizing continuous generative features. This latent grain space is based on an encoder network that models $q_\phi(\mathbf{z}_{i}\vert\mathbf{x}_{i})$ paired with a decoder network $p_\theta(\mathbf{x}_{i}\vert\mathbf{z}_{i})$ allowing to recover $\hat{\mathbf{x}}_{i}$ for every grains $\mathbf{x}_{i}\in\mathcal{X}$. We use the Variational Auto-Encoder \cite{vae} with a mean-field family and Gaussian prior to learn a smooth latent distribution $p(\mathbf{z})$.




\subsection{Latent path encoder}\label{subsec:section_encoder}
As we will perform overlap-add reconstruction, our model processes series of $g$ grains $\mathbf{s}_\mathbf{x}=\{\mathbf{x}_1,\ldots,\mathbf{x}_g\}$ extracted from a given waveform $\mathbf{x}$. The down-sampling ratio between the waveform duration $T$ and number of grains $g$ is given by the hop size separating neighboring grains. Each of these grains $\mathbf{x}_i$ is analyzed separately by the encoder in order to produce $q_\phi(\mathbf{z}_i\vert\mathbf{x}_i)=\mathcal{N}(\bm{\mu}(\mathbf{x}_i),\bm{\sigma}(\mathbf{x}_i))$. Hence, the successive encoded grains form a corresponding series $\mathbf{s}_\mathbf{z}=\{\mathbf{z}_1,\ldots,\mathbf{z}_g\}$ of latent coordinates such that
\begin{equation}
\mathbf{z}_i=\bm{\mu}(\mathbf{x}_i)+\bm{\varepsilon}*\bm{\sigma}(\mathbf{x}_i)
\label{eq:sample_vae}
\end{equation}
with $\bm{\varepsilon}\sim\mathcal{N}(\mathbf{0},\mathbf{I})$. The layers of the encoder are first strided residual convolutions that successively down-sample the input grains through temporal 1-dimensional filters. The output of these layers is then fed into several fully-connected linear layers that map to Gaussian means and variances at the desired latent dimensionality $d_{z}$.

\subsection{Spectral filtering decoder}\label{subsec:section_decoder}
Given a latent series $\mathbf{s}_\mathbf{z}$, the decoder must first synthesize each grain prior to the overlap-add operation. To that end, we introduce a filtering model that adapts the design of \cite{ddsp} to granular synthesis. Hence, each $\mathbf{z}_i$ is processed by a set of residual fully-connected layers that produces frequency domain coefficients $\mathbf{H}_i\in\mathbb{R}^{d_{h}}$ of a filtering module that transforms uniform noise excitations $\mathbf{n}_i\sim\mathcal{U}_{[-1,1]}^{d_{x}}$ into waveform grains. We replace the recurrence over envelope features proposed in \cite{ddsp} by performing separate forward passes over overlapping grain features. Denoting the Discrete Fourier Transform $\text{DFT}$ and its inverse $\text{iDFT}$, this amounts to computing 
\begin{align}
    \hat{\mathbf{X}}_i &= \mathbf{H}_i*\text{DFT}(\mathbf{n}_i) \\
    \hat{\mathbf{x}}_i &= \text{iDFT}(\hat{\mathbf{X}}_i).
\end{align}
Since the DFT of a real valued signal is Hermitian, symmetry implies that for an even grain size $d_{x}$, the network only filters the $d_{h}=d_{x}/2+1$ positive frequencies.

These grains are then used in an overlap-add mechanism that produces the waveform, which is passed through a final learnable post-processing inspired from \cite{wavegan}. This module applies a multi-channel temporal convolution that learns a parallel set of time-invariant FIR filters and improves the audio quality of the assembled signal $\hat{\mathbf{x}}$.

\subsection{Sequence trajectories  embedding}\label{subsec:section_sequence_embedding}
As argued earlier, generative audio models need to sample audio events with a consistent long-term temporal structure. Our model provides this in an efficient manner, by learning a higher-level distribution of sequences $s_\psi(\mathbf{s}_\mathbf{z})$ that models temporal trajectories in the granular latent space $\mathbf{s}_\mathbf{z}\in\mathbb{R}^{d_{z}*g}$. This allows to use the down-sampling of an intermediate frame-level representation in order to learn longer-term relationships. This is achieved by training a temporal recurrent neural network on ordered sequences of grain features $\mathbf{s}_\mathbf{z}$. This process can be applied equivalently to any types of audio signals. As a result, our proposal can also synthesize and transfer meaningful temporal paths inside the latent grain space. It starts by sampling $\mathbf{e}\in\mathbb{R}
^{d_{e}}$ from the Gaussian $\mathbf{e}\sim \mathcal{N}(\mathbf{0},\mathbf{I})$, then sequentially decoding
$s_\psi(\mathbf{s}_\mathbf{z}\vert\mathbf{e})$ and finally generating the grains and overlap-add waveform with $p_\theta(\hat{\mathbf{x}}\vert\mathbf{s}_\mathbf{z})$.

\subsection{Multi-scale training objective}\label{subsec:section_multi_scale}
To optimize the waveform reconstruction, we rely on a multi-scale spectrogram loss \cite{nsf,ddsp}, where STFTs are computed with increasing hop and window sizes, so that the temporal scale is down-sampled while the spectral accuracy is refined. We use both linear and log-frequency STFT \cite{nnaudio} on which we compare log-magnitudes $l(\mathbf{x})=\log(\epsilon+|\text{STFT}[\mathbf{x}]|^2)$ with the L1 distance $||.||_1$. In addition to fitting multiple resolutions of $\text{STFT}_{1\ldots N}$, we can explicitly control the trade-off between low and high-energy components with the $\epsilon$ floor value \cite{sing}. In order to optimize a latent grain space, KL regularization and sampling (\ref{eq:sample_vae}) are performed for each latent point $\mathbf{z}_i$, thus we extend the original VAE objective (\ref{eq:ELBO}) as
\begin{equation}
\resizebox{.9\hsize}{!}{$
\mathcal{L}_{\theta,\phi}=\underbrace{\sum_{n=1}^N||l_n(\mathbf{x})-l_n(\hat{\mathbf{x}})||_1}_{\text{reconstructions}}+\beta*\underbrace{\sum_{i=1}^g \mathcal{D}_{KL} \big[ q_\phi(\mathbf{z}_i|\mathbf{x}_i) \parallel p_\theta(\mathbf{z}) \big]}_{\text{regularizations}}$}
\label{eq:GANG_loss}
\end{equation}
where $N$ is the number of scales in the spectrogram loss and $g$ is the number of grains processed in one sequence.

\section{Experiments}\label{sec:section_exp}

\subsection{Datasets}\label{subsec:datasets}
In order to evaluate our model across a wide variety of sound domains, we train on the following datasets
\begin{enumerate}
\item \textit{Studio-On-Line} provides individual note recordings sampled at 22050 Hz with labels (pitch, instrument, playing technique) for 12 orchestral instruments. The tessitura for \textit{Alto-Saxophone, Bassoon, Clarinet, Flute, Oboe, English-Horn,  French-Horn,  Trombone,  Trumpet, Cello,  Violin, Piano} are in average played in 10 different extended techniques. The full set amounts to around 15000 notes \cite{sol}.
\item \textit{8 Drums} around 6000 one-shot recordings sampled at 16000 Hz in \textit{Clap, Cowbell, Crash, Hat, Kick, Ride, Snare, Tom} instrument classes\footnote{\url{https://github.com/chrisdonahue/wavegan/tree/v1}}.
\item \textit{10 animals} contains around 3 minutes of recordings sampled at 22050 Hz for each of \textit{Cat, Chirping Birds, Cow, Crow, Dog, Frog, Hen, Pig, Rooster, Sheep} classes of the ESC-50 dataset\footnote{\url{https://github.com/karolpiczak/ESC-50}}.
\end{enumerate}
For datasets sampled at 22050 Hz, we use a grain size $d_{x}=2048$, which subsequently sets the filter size $d_{h}=1025$, and compute spectral losses for STFT window sizes $[128, 256, 512, 1024, 2048]$. For datasets sampled at 16000 Hz, $d_{x}=1024$ and STFT window sizes range from 32 to 1024. Hop sizes for both grain series and STFTs are set with an overlap ratio of 75\%. Log-magnitudes are computed with a floor value $\epsilon=5e^{-3}$. Dimensions for latent features are $d_{z}=96$ and  $d_{e}=256$.

\subsection{Models}\label{subsec:model_variants}
Since datasets provide some labels, we both train unconditional models and variants with decoder conditioning. For instance \textit{Studio-On-Line} can be trained with control over pitch and/or instrument classes when using multiple instrument subsets. Otherwise for a single instrument we can instead condition on its playing styles (such as \textit{Pizzicato} or \textit{Tremolo} for the \textit{violin}). To do so, we concatenate one-hot encoded labels $\mathbf{oh_{\text{class}}}$ to the latent vectors at the input of the decoder. During generation we can explicitly set these target conditions, which provide independent controls over the considered sound attributes
\begin{equation}
p_\theta : (\mathbf{s}_\mathbf{z},\mathbf{oh_{\text{class}}})\xrightarrow{}\hat{\mathbf{s}}_\mathbf{x}^{\text{cond.}}\xrightarrow{}\hat{\mathbf{x}}^{\text{cond.}}.
\label{eq:conditioning}
\end{equation}

\subsection{Training}\label{subsec:section_training}
The model is trained according to eq. \ref{eq:GANG_loss}. In the first epochs only the reconstruction is optimized, which amounts to $\beta=0$. This regularization strength is then linearly increased to its target value, during some warm-up epochs. The last epochs of training optimize the full objective at the target regularization strength, which is roughly fixed in order to balance the gradient magnitudes when individually back-propagating each term of the objective. The number of training iterations vary depending on the datasets, we use a minibatch size of 40 grain sequences, an initial learning rate of $2e^{-4}$ and the ADAM optimizer. In this setting, a model can be fitted within 10 hours on a single GPU, such as an Nvidia Titan V.

\section{Results}\label{sec:section_results}
The model performance is first compared to some baseline auto-encoders in Table \ref{tab:baseline}. To assess the generative qualities of the model, we provide audio samples of data reconstructions as well as examples of neural granular sound synthesis\footnote{\url{https://adrienchaton.github.io/neural_granular_synthesis/} \label{url_ano}}. These are generations based on its common processes as well as novel interactions enabled by our proposed neural architecture.

\subsection{Baseline comparison}\label{subsec:baselin_comp}
In the first place, the granular VAE could be implemented using a convolutional decoder that symmetrically reverts the latent mapping of the encoder we use. Strided down-sampling convolutions can be mirrored with transposed convolutions or up-sampling followed with convolutions. We will refer to these baselines as $\text{VAE}_{tr}$ and $\text{VAE}_{up}$ while our model with spectral filtering decoder is $\text{VAE}_{fi}$ and with the added learnable post-processing is $\text{VAE}_{fi+pp}$. We train these models on the \textit{Studio-On-Line} dataset for the full orchestra in \textit{ordinario} and the \textit{strings} in all playing modes as well as the \textit{8 Drums} dataset, keeping all other hyper-parameters identical. We report their test set spectrogram reconstruction scores for the Root Mean Squared Error (RMSE), Log-Spectral Distance (LSD) and their average time per training iteration. Each model was trained for about 10 hours. Accordingly, we can see that our proposal globally outperforms the convolutional decoder baselines, while training and generating fast. The latency of our model to synthesize 1 second of audio is about 19.7 ms. on GPU and 25.0 ms. on CPU.
\begin{table}[ht]
\begin{center}
\resizebox{0.9\hsize}{!}{ 
\begin{tabular}{c|c|c|c|c}
\cline{2-5}
                                  \multicolumn{1}{c}{} & \multicolumn{1}{c|}{$\text{\textbf{VAE}}_{\mathbf{tr}}$} & $\text{\textbf{VAE}}_{\mathbf{up}}$ & $\text{\textbf{VAE}}_{\mathbf{fi}}$ & $\text{\textbf{VAE}}_{\mathbf{fi+pp}}$ \\ \specialrule{.2em}{0em}{0em}
\multicolumn{5}{c}{\textit{\textit{Studio-On-Line ordinario}}}                                                    \\ \hline
\multicolumn{1}{c|}{RMSE}        &         6.86          &         6.65          &  6.22  &      \textbf{4.86}          \\ \hline
\multicolumn{1}{c|}{LSD}         &         1.60          &         1.62          &  1.29  &      \textbf{1.17}          \\ \specialrule{.2em}{0em}{0em}
\multicolumn{5}{c}{\textit{\textit{Studio-On-Line strings}}}                                                    \\ \hline
\multicolumn{1}{c|}{RMSE}        &         5.68          &         5.78          &  5.29  &      \textbf{4.07}          \\ \hline
\multicolumn{1}{c|}{LSD}         &         1.39          &         1.43          &  1.19  &      \textbf{1.05}          \\ \specialrule{.2em}{0em}{0em}
\multicolumn{5}{c}{\textit{8 Drums}}                                                  \\ \hline
\multicolumn{1}{c|}{RMSE}        &         3.85          &         4.39          &  \textbf{2.65}  &      2.79          \\ \hline
\multicolumn{1}{c|}{LSD}         &         0.94          &         0.66          &  \textbf{0.52}  &      \textbf{0.52}          \\ \specialrule{.2em}{0em}{0em}
\multicolumn{1}{c|}{sec./iter}    &        2.32          &         2.87           & \textbf{0.54}  &   0.58             \\ \hline
\end{tabular}
}
\end{center}
\caption{Report of the baseline model comparison. Bold denotes the best model for each evaluation.}
\label{tab:baseline}
\end{table}

\subsection{Common granular synthesis processes}\label{subsec:common_inter}
The audio-quality of the models trained in different sound domains can be judged by data reconstructions. It gives a sense of the model performance at auto-encoding various types of sounds. This extends to generating new sounds by sampling latent sequences rather than encoding features from input sounds. For structured one-shot samples, such as musical notes and drum hits, latent sequences are generated from the higher-level sequence embedding. For use in composition (e.g. MIDI score), this sampling can be done with conditioning over user classes such as pitch and target instrument (eq. \ref{eq:conditioning}). Since the VAE learns a continuously invertible grain space, it can as well be explored with smooth interpolations that render free-synthesis trajectories. Some multidimensional latent curves that are mapped to overlap-add grain sequences, including linear interpolations between random samples from the latent Gaussian prior, circular paths and spirals. When repeating forward and backward traversals of a linear interpolation or looping a circular curve, we can modulate non-uniformly the steps between latent points in order to bring additional expressivity to the synthesis. Free-synthesis can be performed at variable lengths (in multiples of $g$) by concatenating several contiguous latent paths.

\subsection{Audio style and temporal manipulations}\label{subsec:audio_style}
\begin{figure}[ht]
\centering
\includegraphics[width=0.9\columnwidth]{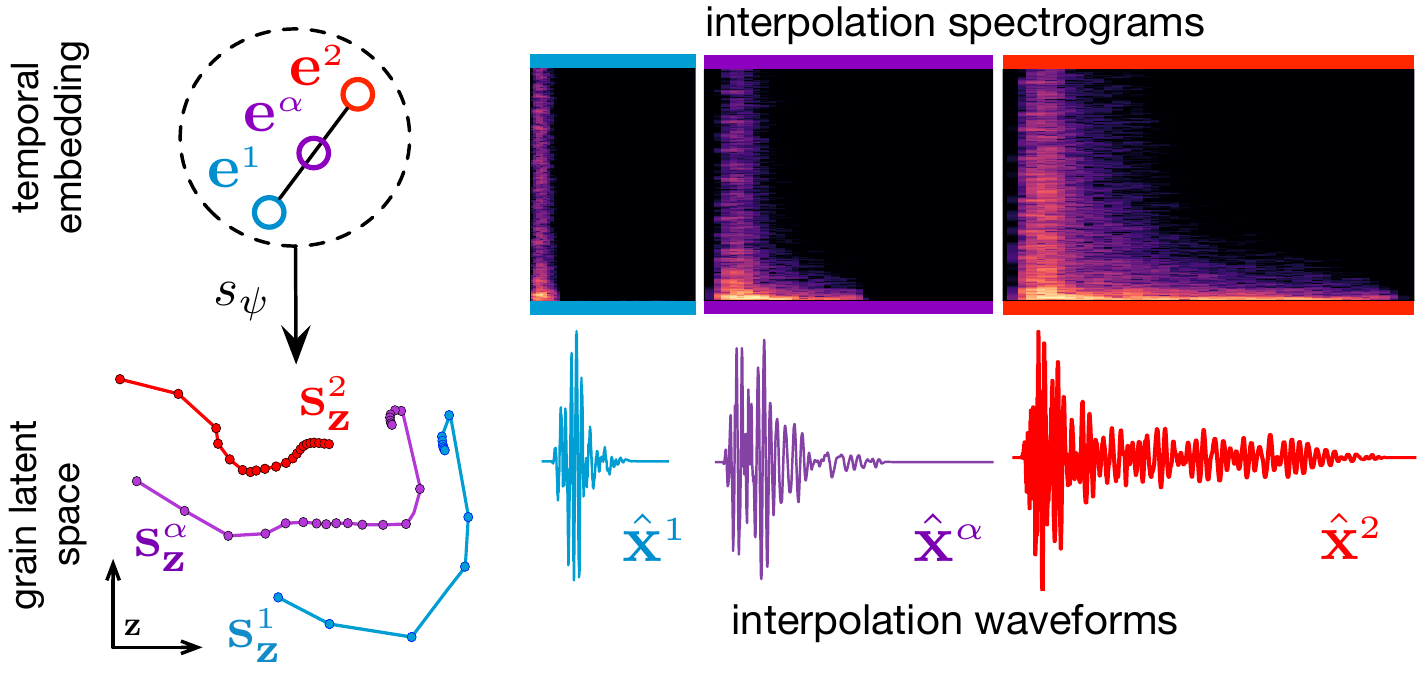}
\caption{An interpolation in the continuous temporal embedding generates series of latent grain features corresponding to waveforms with evolving temporal structure. Here three drum sounds with increasingly sustained envelope. The point $\mathbf{e}^{_\alpha}$ is set half-way from $\mathbf{e}^{_1}$ and $\mathbf{e}^{_2}$.\label{fig:e_interp}}
\end{figure}

To perform data-driven resynthesis, a target sample is analyzed by the encoder. Its corresponding latent features are then decoded, thus emulating the target sound in the \textit{style} of the learned grain space. A conditioning over multiple \textit{timbres} (e.g. instrument classes) allows for finer control over such audio transfer between multiple target \textit{styles}. To perform resynthesis of audio samples longer than the grain series length $g$, we auto-encode several contiguous segments that are assembled with fade-out/fade-in overlaps. Since the model can also learn a continuous temporal embedding, by interpolating this higher-level space, we can generate successive latent series in the grain space that are decoded into signals with evolving temporal structures. We illustrate this feature in Figure \ref{fig:e_interp}.

\subsection{Real-time sound synthesis}\label{subsec:implementation}
With GPU support, for instance a sufficient dedicated laptop chip or an external thunderbolt hardware, the models can be ran in real-time. In order to apply trained models to these different generative tasks, we currently work on some prototype interfaces based on a \textit{Python OSC}\footnote{\url{https://pypi.org/project/python-osc/}} server controlled from a \textit{MaxMsp}\footnote{\url{https://cycling74.com}} patch. For instance a neural drum machine \textsuperscript{\ref{url_ano}} featuring a step-sequencer driving a model with sequential embedding and conditioning trained over the \textit{8  Drums} dataset classes.

\section{Conclusions}\label{sec:conclusion}
We propose a novel method for raw waveform generation that implements concepts from granular sound synthesis and digital signal processing into a Variational Auto-Encoder. It adapts to a variety of sound domains and supports neural audio modeling at multiple temporal scales. The architecture components are interpretable with respect to its spectral reconstruction power. Such VAE addresses some limitations of traditional techniques by learning a continuously invertible grain latent space and a hierarchical temporal embedding. Moreover, it enables multiple modes of generation derived from granular sound synthesis, as well as potential controls for composition purpose. By doing so, we hope to enrich the creative use of neural networks in the field of musical sound synthesis.

\begin{acknowledgments}
This JSPS International Research Fellow research was partially supported by JST AIP Acceleration Research Grant Number JPMJCR20U3, as well as ANR:17-CE38-0015-01 MAKIMOno project, SSHRC:895-2018-1023 ACTOR Partnership, Emergence(s) ACIDITEAM project from Ville de Paris and ACIMO project of Sorbonne Université.
\end{acknowledgments}

\bibliography{icmc2020template}

\end{document}